\documentclass[%
reprint,
 amsmath,amssymb,
 aps,
prb,
floatfix
]{revtex4-2}

\usepackage{graphicx}% Include figure files
\usepackage{dcolumn}% Align table columns on decimal point
\usepackage{bm}% bold math
\usepackage{hyperref}% add hypertext capabilities
\usepackage{enumitem}
\usepackage{physics}

\def\mb#1{\mathbf{#1}}

\begin{document}
\title{Low bending rigidity and large Young's modulus drive strong flexural phonon renormalization in two-dimensional monolayers}
% Predicting kirigami in two-dimensional monolayers from anharmonic renormalization of elastic constants
% Predicting kirigami in two-dimensional monolayers from renormalization of elastic constants with temperature and system size
% Giant renormalization of bending rigidity and Young's modulus of two-dimensional monolayers with temperature and system size
% Anharmonic renormalization of elastic constants in two-dimensional monolayers
% Small curvature of flexural phonon dispersion induces strong anharmonic renormalization of elastic constants in two-dimensional monolayers
% Weak bending rigidity induces strong anharmonic renormalization of elastic constants in two-dimensional monolayers
% Temperature and sample size-dependent modulation of elastic constants of two-dimensional group IV and III-V monolayers
\author{Navaneetha K. Ravichandran}
\email{navaneeth@iisc.ac.in}
\affiliation{%
 Department of Mechanical Engineering, Indian Institute of Science, Bangalore 560012, India
 }%

\date{\today}% It is always \today, today,
% Anharmonic renormalization of elastic constants of two-dimensional semiconductor monolayers

\begin{abstract}
Many intriguing phenomena such as the wave-like hydrodynamic heat flow, the logarithmic divergence of electrical resistivity at low temperatures and microscale kirigami are driven by flexural acoustic (ZA) phonons in two-dimensional (2D) materials. Yet, a definitive first-principles description of their dispersion, with explicit consideration of the crystal anharmonicity and the stability of large 2D monolayers against thermal fluctuations, is lacking in the literature. Using first-principles calculations, we show that the bending rigidity ($\kappa$) controls the anharmonic renormalization of the ZA phonons throughout the Brillouin zone in 2D monolayers, with stronger renormalization in low-$\kappa$ materials like germanene and weaker effects in high-$\kappa$ materials like molybdenum disulphide. Furthermore, the ZA phonons at long wavelengths undergo an additional renormalization to stabilize the flat phase of the 2D monolayers against thermal fluctuations, which is modulated by the competing influence of the bending rigidity and the in-plane Young’s modulus in all materials. The resulting renormalized ZA phonon dispersions are qualitatively and quantitatively different from those commonly used by the first-principles community, thus motivating a re-examination of the ZA phonon-driven unconventional thermal and electronic phenomena in 2D as well as lower-dimensional systems. Our work provides new insights into the role of nanoscale crystal anharmonicity and macroscale elasticity in shaping the vibrational properties of 2D materials and will inform novel engineering applications that are exclusive to low dimensions such as kirigami, with materials beyond graphene. 
\end{abstract}

\maketitle
\clearpage

\section{Introduction}
Flexural acoustic (ZA) phonons, which represent the out-of-plane acoustic vibrational modes of two-dimensional (2D) crystals, are known to play a vital role in their elastic, electronic and thermal properties. For long wavelengths [or small wave vectors ($q$)], the elastic plate theory dictates that the ZA phonons must exhibit a quadratic dispersion given by $\omega_{\text{ZA}} = \sqrt{\frac{\kappa}{\rho_{\text{2D}}}}q^2$, where  $\rho_{\text{2D}}$ is the 2D mass density and $\kappa$ is the bending rigidity – an important elastic property of the 2D materials that impacts novel applications such as kirigami~\cite{blees_graphene_2015}. As dominant heat carriers in 2D materials like graphene, the small-$q$ ZA phonons with a quadratic dispersion are predicted to drive the unconventional hydrodynamic heat flow regime at cryogenic temperatures~\cite{lee_hydrodynamic_2015, li_effects_2025} and to strongly participate in higher-order four-phonon scattering processes that act to suppress the thermal conductivity at room temperature~\cite{han_thermal_2023}. A purely quadratic dispersion of small-$q$ ZA phonons also results in a singularity in the strength of electron-phonon interactions~\cite{fischetti_mermin-wagner_2016}, and the predicted deviations from this perfect quadraticity due to thermal fluctuations~\cite{fasolino_intrinsic_2007, bowick_non-hookean_2017} still leads to anomalous phenomena such as a logarithmically-divergent electrical resistivity of 2D materials at low temperatures ($T$)~\cite{mariani_flexural_2008}.\\

Recognizing its importance, the research community has devoted considerable effort to predict the dispersion relation of ZA phonons accurately using first-principles and atomistic calculations. For example, first-principles calculations at $T \sim$ 0 K have unveiled the importance of enforcing rotational invariance and stress-free equilibrium conditions to obtain purely quadratic dispersions of the ZA phonons in the long-wavelength limit for 2D materials~\cite{carrete_physically_2016, croy_bending_2020, lin_general_2022}. However, a purely quadratic ZA phonon dispersion in 2D materials poses a theoretical difficulty, since the thermal excitation of these low-energy ZA phonons at $T>0$ K destabilizes the long-range atomic ordering in 2D, according to the Hohenberg-Mermin-Wagner theorem~\cite{hohenberg_existence_1967, mermin_absence_1966}. This thermally-driven instability can be resolved by allowing for coupling between the in-plane and the out-of-plane degrees of freedom, which results in a \emph{renormalized}, sub-quadratic ZA phonon dispersion in the small-$q$ limit, as shown using atomistic calculations~\cite{fasolino_intrinsic_2007, los_scaling_2016, bowick_non-hookean_2017} as well as field theoretical approaches~\cite{gazit_correlation_2009, zakharchenko_self-consistent_2010, kosmrlj_response_2016}. These renormalization approaches, however, are not predictive, since they require empirical inputs of the elastic constants of the 2D materials, and do not account for the phonon renormalization throughout the Brillouin zone (BZ), induced by the anharmonic terms of the atomic crystal potential at $T > 0$ K~\cite{ravichandran_unified_2018}. Thus, a single complete first-principles description of the ZA phonon dispersion spanning across the BZ, while accurately accounting for the crystal anharmonicity and stability against thermal fluctuations, that is necessary to predict the novel ZA phonon-driven elastic, thermal and electronic phenomena introduced earlier, is lacking in the literature.\\

Here, we predict, from first-principles, the dispersion relation of ZA phonons in 2D monolayers accounting for the crystal anharmonicity-driven renormalization, while simultaneously including the coupling between the in-plane and the out-of-plane degrees of freedom that protects the long-range ordering in 2D at $T > 0$ K. We find that the $\kappa$ controls the strength of the anharmonic renormalization of the ZA phonons through out the BZ, with the low-$\kappa$ materials exhibiting the strongest renormalization and those with high $\kappa$ exhibiting the weakest effects.  We show that these observations are linked to the large thermal occupation of the ZA phonon branch in low-$\kappa$ materials that drives the strong anharmonic renormalization of the harmonic interatomic force constants (IFC). Furthermore, to stabilize the ordered structure of large-area 2D monolayers, the necessary coupling between the in-plane and the out-of-plane degrees of freedom gives rise to a universal competition between the bending rigidity and the in-plane 2D Young’s modulus ($Y^{\text{2D}}$) towards further renormalizing the small-$q$ ZA phonons, where the former suppresses the renormalization while the latter amplifies the effect in all 2D monolayers. We anticipate that our results will drive future efforts towards revisiting the predictions of unconventional thermal and electronic transport phenomena that relied on the quadraticity of the ZA phonon dispersions and inform the choice of material alternatives to graphene for exciting engineering applications such as kirigami. Given the generality of our first-principles renormalization framework, we expect that our work will inspire exploration of novel elastic, thermal and electronic phenomena in other low-dimensional crystals such as nanotubes and crystalline polymer strands as well.

\section{Results}
\begin{figure*}[!ht]
\centering
\includegraphics[width=\linewidth, trim=4mm 0mm 22mm 0mm, clip]{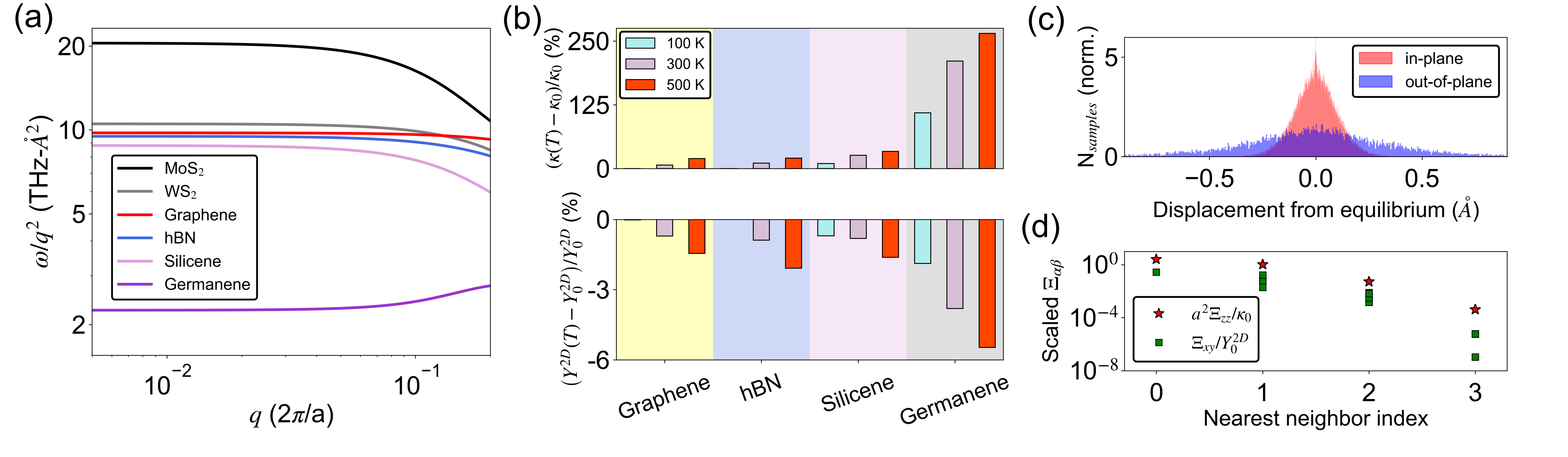}
\caption{\textbf{Origin of strong anharmonic renormalization of ZA phonons and elastic constants in germanene.} \textbf{(a)}. Long wavelength (small-$q$) behavior of $\omega/q^2$ for ZA phonons in different 2D monolayers calculated using the bare harmonic IFCs. The ZA phonons in all 2D monolayers have a quadratic dispersion in the small-$q$ limit, with the curvature $\left[2\omega/q^2\right]$ being the smallest for germanene and the largest for MoS$_2$. \textbf{(b)}. Bar plots of the relative changes in the $\kappa_0\left(T\right)$ and $Y^{\text{2D}}_0\left(T\right)$ of four 2D monolayers caused by the anharmonic renormalization of the bare harmonic IFCs using the SCAP framework. Germanene exhibits the strongest renormalization of the elastic constants due to the small curvature of its ZA phonons at small $q$, resulting in their high thermal occupation that drive large thermal excursions of atoms in the out-of-plane direction, as shown in \textbf{(c)}, thus resulting in strong anharmonic renormalization of the harmonic IFCs. The contrasting relative magnitudes of renormalization of the $\kappa_0\left(T\right)$ and the $Y^{\text{2D}}_0\left(T\right)$ in all materials is caused by the much stronger anharmonic renormalization of the appropriately-scaled cross-plane harmonic IFCs $\left[a^2\Xi_{zz}/\kappa_0 = a^2\left(\psi_{zz} - \chi_{zz}\right)/\kappa_0\right]$ compared to that of their in-plane counterparts $\left[\Xi_{xy}/Y^{\text{2D}}_0\right]$, as shown e.g., for germanene at 300 K in \textbf{(d)}.}\label{fig:Figure1}
\end{figure*}

We compute the crystal anharmonicity-driven renormalization of the ZA phonon dispersions from first-principles using the self-consistent anharmonic phonon (SCAP) framework~\cite{ravichandran_unified_2018} without any adjustable parameters at different temperatures. The bare and the renormalized ZA phonon dispersions are obtained from the bare and the renormalized harmonic IFCs respectively in SCAP, with the strict enforcement of the crystal space group symmetries, translational and rotational invariance and stress-free equilibrium conditions as detailed in the Appendix~\ref{sec:methods_SCAP}. The resulting ZA phonon dispersions are perfectly quadratic in the small-$q$ limit for all 2D monolayers, as seen in Fig.~\ref{fig:Figure1} (a), thus enabling their parametrization in terms of the $\kappa$ of the material as $\omega_{\text{ZA}} = \sqrt{\kappa/\rho_{\text{2D}}}\ q^2$, where $\rho_{\text{2D}}$ represents the 2D mass density~\cite{landau_theory_1986}. The translational invariance condition on the harmonic IFCs results in a linear dispersion in the small-$q$ limit for the in-plane transverse (TA) and longitudinal (LA) acoustic phonons as shown in the Supplementary Figs. 4 and 5, thus allowing their parametrization in terms of the 2D Young's modulus and the Poisson's ratio ($\sigma$) as: $\omega_{\text{LA}} = \sqrt{Y^{\text{2D}}/\left[\rho^{\text{2D}}\left(1-\sigma^2\right)\right]}\ q$ and $\omega_{\text{TA}} = \sqrt{Y^{\text{2D}}/\left[\rho^{\text{2D}}\left(1+\sigma\right)\right]}\ q$~\cite{landau_theory_1986}. These parameterizations not only connect the renormalized ZA phonon dispersions to experimentally-accessible observables such as the $\kappa$ and $Y^{\text{2D}}$, but also provide simple intuitive understanding of the origin of the renormalization phenomenon, particularly for the long-wavelength ZA phonons in 2D, as we show below.

\subsection{Strong anharmonic renormalization of ZA phonons in low-$\kappa$ monolayers}
Figure~\ref{fig:Figure1} (b) shows the relative change in the $\kappa$ and the $Y^{\text{2D}}$ of graphene, hexagonal boron nitride (hBN), silicene and germanene monolayers at different temperatures [$\kappa_0\left(T\right)$ and $Y^{\text{2D}}_0\left(T\right)$], with respect to the corresponding bare values [$\kappa_0$ and $Y^{\text{2D}}_0$], arranged in the decreasing order of $\kappa_0$. These elastic constants are obtained by fitting the renormalized and bare ZA phonon dispersion calculated using the SCAP framework in the small-$q$ limit to the parameterizations introduced earlier. The relative changes in the $\kappa$'s of all of the transition metal dichalcogenides (TMDCs) are comparable to or weaker than that of graphene, so they are not shown here. We see from Fig.~\ref{fig:Figure1} (b) that the $\kappa_0$'s increase and the $Y^{\text{2D}}_0$'s decrease with temperature in all four monolayers. We note that the observed increasing trend in the $\kappa_0\left(T\right)$ for graphene is consistent with the recent experimental measurements~\cite{tomterud_observation_2025} and the magnitudes compare reasonably well with each other.\\ 

From Fig.~\ref{fig:Figure1} (b), we observe two striking features. First, the relative magnitudes of anharmonic renormalization of both elastic constants [$\kappa_0$ and $Y^{\text{2D}}_0$] increase with decreasing $\kappa_0$ as we go from graphene to germanium, with germanium exhibiting significantly stronger renormalization with increasing temperature compared to the other 2D monolayers. This feature is caused by the low $\kappa_0$ of germanene, which results in a large thermal population of the ZA phonons, thus driving large thermal displacements of the germanium atoms in the out-of-plane direction, as shown in Fig.~\ref{fig:Figure1} (c). Such large out-of-plane displacements from equilibrium subject the germanium atoms to an effective renormalized harmonic potential that includes much larger weighted contributions from the quartic interatomic force constants, as described in the Appendix~\ref{sec:methods_SCAP}. Thus, the $\kappa_0$, and consequently, the ZA phonons at long wavelengths, undergo much stronger renormalization in germanene than in the rest of the 2D monolayers. In fact, the curvature of the ZA phonon dispersion of germanene at room temperature exceeds its value at absolute zero by $76\%$ due to the SCAP renormalization. \\

\begin{figure}[!ht]
\centering
\includegraphics[width=\linewidth, trim=0mm 0mm 10mm 0mm, clip]{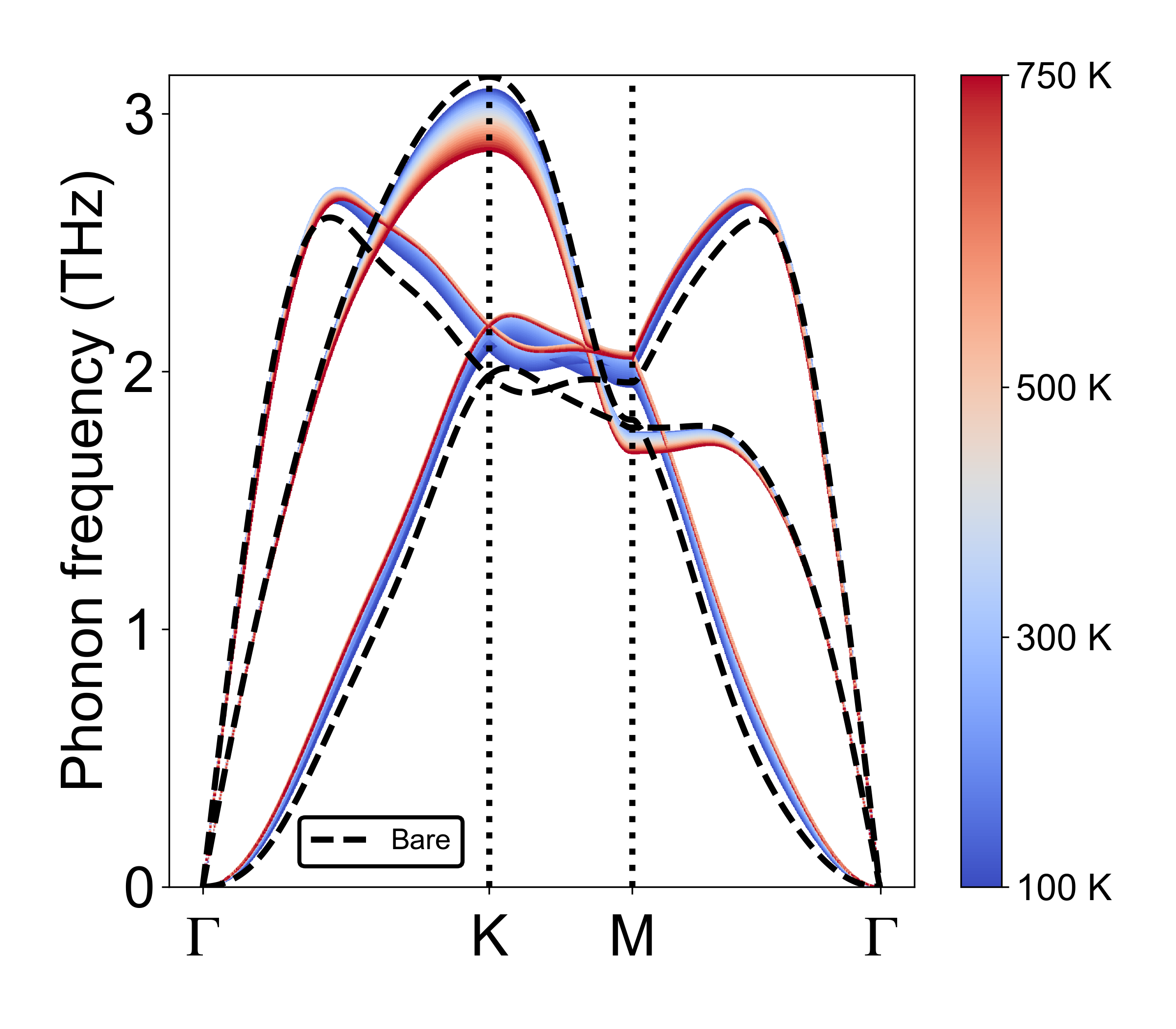}
\caption{\textbf{Temperature-dependent acoustic phonon dispersions in germanene along high symmetry directions, along with the bare (i.e., $T \sim 0$ K) dispersions.} We observe a stronger anharmonic renormalization of ZA phonons relative to that of LA and TA phonons, from the BZ center [$\Gamma$-point] extending till about mid-way to the BZ edge, driven by the small $\kappa_0$ of germanene.}  \label{fig:Figure2}
\end{figure}

Second, the relative change in $\kappa_0$ with temperature obtained with the SCAP renormalization is much larger than that for $Y^{\text{2D}}_0$ for all 2D monolayers, indicating a stronger effect of the anharmonic renormalization on the out-of-plane ZA phonons than on the in-plane LA and TA phonon branches, as is evident in Fig.~\ref{fig:Figure2} for the zone-center acoustic phonons in germanene. The root cause of this difference in the strength of anharmonic renormalization of the in-plane and the out-of-plane polarized phonons is the significant difference between the thermal populations of the three acoustic phonon branches within each material. Owing to their lower vibrational frequencies, the ZA phonons that determine $\kappa_0$ have a significantly larger thermal population than the in-plane LA and TA branches that determine $Y^{\text{2D}}$ at all temperatures. Thus, all 2D materials exhibit significantly larger out-of-plane displacements relative to those within the plane, as shown, e.g., in Fig.~\ref{fig:Figure1} (c) for germanene at 300 K. Therefore, the out-of-plane fluctuations encounter a strongly-renormalized harmonic potential environment compared to the in-plane vibrations on an average, which translates into a much stronger relative anharmonic renormalization of $\kappa_0$ compared to $Y^{\text{2D}}$ in all 2D monolayers. \\

These qualitative explanations for the strength of anharmonic renormalization of the elastic constants, and therefore, the phonon dispersions at long wavelengths, can be precisely quantified by a perturbative analysis of the SCAP renormalization procedure at small-$q$. As derived in the Supplementary Section 3, the relative change in $\kappa_0$ after renormalization takes the form:
\begin{align}
    &\frac{\Delta \kappa_0}{\kappa_0} = \frac{\kappa_0\left(T\right) - \kappa_0}{\kappa_0} \sim \frac{a^2\Xi_{zz}^{\left(M\mu, N\nu\right)}}{\kappa_0} \label{eq:kappa_scaling}
\end{align}
where $\Xi^{\left(M\mu, N\nu\right)}_{\alpha\beta}$ is the difference between the renormalized and the bare components of harmonic IFCs between the pair of atoms at the lattice sites $M$ and $N$ and basis sites $\mu$ and $\nu$ respectively, with $\alpha$ and $\beta$ being the Cartesian components of the IFC tensors. Similarly, the relative change in $Y^{\text{2D}}_0$ after renormalization takes the form:
\begin{align}
\frac{\Delta Y^{\text{2D}}_0}{Y^{\text{2D}}_0} &= \frac{Y_0^{\text{2D}}\left(T\right) - Y_0^{\text{2D}}}{Y_0^{\text{2D}}}\sim \frac{\Xi_{\alpha\beta}^{\left(M\mu, N\nu\right)}}{Y_0^{\text{2D}}}  \label{eq:Y_scaling}
\end{align}
where $\alpha, \beta \ne z$. We find from Table~\ref{tab:params_dispersion} that the estimate of $\frac{\Delta \kappa_0}{\kappa_0}$, obtained by considering the maximum values of $\Xi^{\left(M\mu, N\nu\right)}_{\alpha\beta}$, is by far the largest for the material with the smallest $\kappa_0$ - germanene. Furthermore, the estimates of $\frac{\Delta \kappa_0}{\kappa_0}$ far exceed those of $\frac{\Delta Y^{\text{2D}}_0}{Y^{\text{2D}}_0}$ for all materials. Finally, these trends apply not just to the maximum values of $\Xi^{\left(M\mu, N\nu\right)}_{\alpha\beta}$, but also to the individual components of the $\Xi$ tensors for different nearest neighbors, as shown in Fig.~\ref{fig:Figure1} (d) for germanene at 300 K. Although this perturbative treatment about $q=0$ provides an understanding of the ZA phonon renormalization at long wavelengths, the temperature-dependence of the phonon dispersions closer to the edge of the Brillouin zone is more complex, as shown in Fig.~\ref{fig:Figure2}, which must therefore be obtained from a rigorous calculation using the SCAP framework, as detailed in the Appendix~\ref{sec:methods_SCAP}.

\renewcommand{\arraystretch}{2}
\begin{table}[h]
\centering
\begin{tabular}{cccccccc}
\hline
Mat. & $\kappa_0$ & $Y^{\text{2D}}_0$ & $a^2\left|\Xi_{zz}^{\text{max}}\right|$ & $\left|\Xi_{\alpha\beta}^{\text{max}}\right|$ & $\left(\frac{\Delta \kappa_0}{\kappa_0}\right)$ & $\left(\frac{\Delta Y^{\text{2D}}_0}{Y^{\text{2D}}_0}\right)$\\
& [eV] & [eV/\AA$^2$] & [eV] & [eV/\AA$^2$] & [Est.] & [Est.]\\
\hline
C & 1.81 & 22.01 &  0.3 & 1.13 & 0.17 & 0.05\\
hBN & 1.7 & 14.82 & 0.65 & 0.71 & 0.38 & 0.05\\
Si & 1.36 & 3.98 & 2.51 & 0.88 & 1.85 & 0.22\\
Ge & 0.23 & 3.26 & 1.8 & 0.87 & 7.83 & 0.27\\
MoS$_2$ & 29.67 & 6.81 & 1.07 & 0.19 & 0.04 & 0.03\\
WS$_2$ & 12.55 & 7.96 & 2.0 & 0.2 & 0.16 & 0.03\\
\hline
\end{tabular}
\caption{\textbf{Parameters describing the small-$q$ acoustic phonon dispersions for different 2D monolayers.} In the fourth and the fifth columns, the largest magnitudes of the $zz$ and $\alpha(\ne z)\beta(\ne z)$ components of $\Xi_{\alpha\beta}^{\left(M\mu, N\nu\right)}$ at 300 K are used. In the last two columns, estimates of the relative change of $\kappa_0\left(T\right)$ and $Y^{\text{2D}}_0\left(T\right)$ at 300 K with respect to the corresponding bare values - $\kappa_0$ and $Y^{\text{2D}}_0$ are also reported for each monolayer, following Eqs.~\ref{eq:kappa_scaling} and~\ref{eq:Y_scaling}, with replacement of $\Xi_{\alpha\beta}$ by $\left|\Xi_{\alpha\beta}^{\text{max}}\right|$. These estimates capture the trends of significantly large relative enhancements of $\kappa_0$'s and $Y^{\text{2D}}_0$'s with temperature for all 2D monolayers observed in Fig.~\ref{fig:Figure1} (b).} \label{tab:params_dispersion}
\end{table}

\subsection{Renormalization of small-$q$ ZA phonons to stabilize large 2D monolayers}
The renormalized ZA phonons with quadratic dispersions at small $q$ and the resulting temperature-dependent elastic constants [$\kappa_0\left(T\right)$ and $Y^{\text{2D}}_0\left(T\right)$] are applicable for small suspended 2D monolayers ($<$ 100 atoms~\cite{los_scaling_2016}) as well as for the 2D samples supported on a weakly-interacting substrate~\cite{amorim_flexural_2013, tomterud_observation_2025}. For larger samples, these quadratic ZA phonon dispersions at long wavelengths pose a fundamental difficulty. Quadratic dispersion of the out-of-plane ZA phonons for a stress-free 2D monolayer of area $A$ in the long-wavelength (elastic) limit arises from the minimization of the bending free energy of the form: $F\left(h\right) = \frac{1}{2A}\int \kappa_0\left(\nabla^2h\right)^2\mathrm{d}\mb{x}$, where $h\left(\mb{x}\right)$ represents the out-of-plane displacement fluctuations at point $\mb{x}$ in the plane of the monolayer~\cite{landau_theory_1986}. As shown in Ref.~\cite{nelson_membraneStatMech_2004}, for this free energy, the correlation function of the angles [$\theta\left(x, y\right)$] made by the local normals with the out-of-plane direction takes the form: $\expval{\left[\theta\left(x, y\right)\right]^2} \approx k_B T\int \frac{\mathrm{d}^2q}{\left(2\pi\right)^2}\frac{1}{\kappa_0 q^2}$. For a constant $\kappa_0$, this correlation function becomes logarithmically divergent with the system size $L\approx 2\pi/q$, taking the form: $\expval{\left[\theta\left(x, y\right)\right]^2} \approx \frac{k_B T}{\kappa}\ln\left(L/a\right)$, where $a$ is the lattice spacing. Thus, a perfectly quadratic ZA phonon dispersion forbids long-range 2D structural ordering present in the flat phase of a 2D crystalline monolayer, which is a consequence of the Hohenberg-Mermin-Wagner theorem~\cite{hohenberg_existence_1967, mermin_absence_1966}. \\

However, recent works on the atomistic and field-theoretical description of 2D membranes have shown that, in reality, low energy out-of-plane thermal fluctuations couple with the in-plane degrees of freedom through the higher-order terms in the free energy~\cite{aronovitz_fluctuations_1988, le_doussal_self-consistent_1992}. The consequence of this coupling is a system size-dependent $\kappa$ that restores the long-range order in the flat phase of the 2D monolayers~\cite{nelson_fluctuations_1987, nelson_membraneStatMech_2004}. These calculations involving the renormalization group theory (RG)~\cite{aronovitz_fluctuations_1988, kosmrlj_response_2016}, self-consistent screening approximation (SCSA)~\cite{le_doussal_self-consistent_1992, gazit_structure_2009, zakharchenko_self-consistent_2010, kosmrlj_mechanical_2013} and atomistic simulations~\cite{roldan_suppression_2011, los_scaling_2009, los_scaling_2016, bowick_non-hookean_2017} have been performed in the past with the primary goal of obtaining a scaling relationship for an effective $\kappa$ which depends on the size of the 2D system $\left[\kappa\left(L\right)\right]$ in the form of a power law: $\kappa\left(L\right)\sim L^{\eta_\kappa}$ for large $L$, with the power-law exponent, $\eta_\kappa$, predicted to lie between $0.82-0.85$. However, these calculations have been limited to graphene only and require empirical inputs such as the interatomic potentials (for the atomistic calculations) and the values of $\kappa_0\left(T\right)$ and $Y^{\text{2D}}_0\left(T\right)$ (for the RG and the SCSA calculations).\\

We calculate $\kappa\left(L\right)$ by solving the SCSA equations from first-principles, as detailed in the Appendix~\ref{sec:methods_SCSA}, using the $\kappa_0\left(T\right)$ and $Y^{\text{2D}}_0\left(T\right)$ predicted with the SCAP framework earlier for different 2D monolayers. In the SCSA framework, the renormalization of $\kappa$ with the sample size $L \approx 2\pi/q$ is calculated by solving the Dyson's equation, $G^{-1}\left(\mb{q}, T\right) = G_0^{-1}\left(\mb{q}, T\right) + \Sigma\left(\mb{q}, T\right)$ for the renormalized propagator $G^{-1}\left(\mb{q}, T\right) = \frac{\kappa\left(\mb{q}, T\right)}{k_B T}q^4$ in a self-consistent manner. Here, $G_0^{-1}\left(\mb{q}, T\right) = \frac{\kappa_0\left(T\right)}{k_B T}q^4$ is the bare propagator at the temperature $T$ and $\Sigma\left(\mb{q}, T\right) = 2\int\frac{d^2\mb{k}}{\left(2\pi\right)^2}\frac{b_0\left(T\right)}{1 + b_0\left(T\right)I\left(\mb{k}\right)}\|\hat{\mb{k}}\times\mb{q}\|_2^4G\left(\mb{q}-\mb{k}\right)$ is the self energy of the interaction with $b_0\left(T\right) = \frac{Y^{\text{2D}}_0\left(T\right)}{2k_B T}$.

\begin{figure*}[!ht]
\centering
\includegraphics[width=\linewidth, trim=4mm 0mm 26mm 0mm, clip]{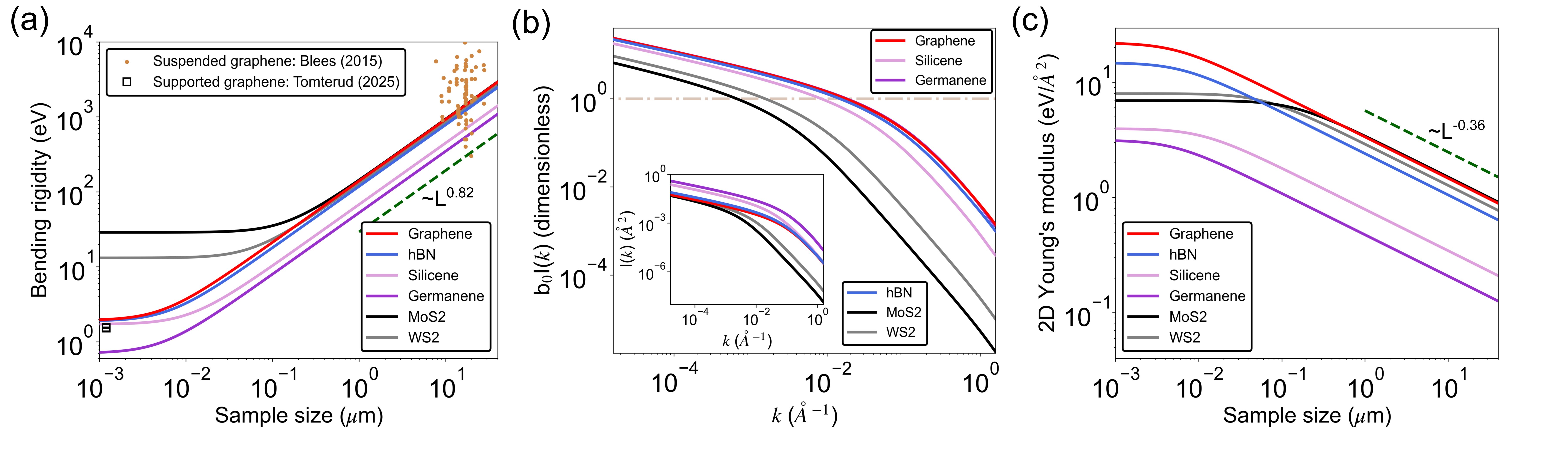}
\caption{\textbf{Renormalization of the elastic constants of 2D monolayers with system size at 300 K.} \textbf{(a)} Bending rigidity, $\kappa$, of 2D monolayers as a function of the sample size ($L$), showing a universal scaling of $\kappa\left(L\right)\sim L^{0.82}$ for samples larger than a micron in all 2D monolayers while approaching their respective $\kappa_0\left(T\right)$ for nanoscale samples. The calculated $\kappa$'s agree well with the available measurements in the literature for graphene that span several orders of magnitude from $\sim 1$ eV for a supported sample that mimics a nanoscale system to $\sim 10^3$ eV for suspended samples with dimensions exceeding 10 $\mu$m. For the suspended samples, the system size is chosen as a geometric mean of the reported lengths and widths in Ref.~\cite{blees_graphene_2015}. \textbf{(b)} Dimensionless vacuum polarization [$b_0\left(T\right)I\left(k\right)$] as a function of the wave vector $k$ for different 2D materials, which causes the transition from the small-$L$ [$b_0\left(T\right)I\left(k\right) \ll 1$] to large-$L$ [$b_0\left(T\right)I\left(k\right) \gg 1$] trends in the elastic constants through the SCSA equations. The quantity $I\left(k\right)$ is plotted as an inset for reference. \textbf{(c)} Two-dimensional Young's modulus, $Y^{\text{2D}}$, of 2D monolayers as a function of $L$, showing a weaker universal scaling relative to that of $\kappa$ $\left[Y^{\text{2D}}\left(L\right)\sim L^{-0.36}\right]$ for samples larger than a micron in all 2D monolayers, while approaching their respective $Y^{\text{2D}}_0\left(T\right)$ for nanoscale samples. The scaling exponents of $\kappa$ $\left[\eta_\kappa\right]$ and $Y^{\text{2D}}$ $\left[\eta_Y\right]$ satisfy the Ward identity for the 2D rotation group given by $\eta_Y = 2\eta_\kappa - 2$.}  \label{fig:Figure3}
\end{figure*}

\subsection{Competing influences of $\kappa$ and $Y^{\text{2D}}$ in renormalizing small-$q$ ZA phonons}
Figure~\ref{fig:Figure3} (a) shows the $\kappa\left(L\right)$ for different 2D monolayers considered in this work, calculated using the SCSA at 300 K. The predicted $\kappa\left(L\right)$ agree well with the experimental measurements on supported graphene in the presence of a weakly interacting substrate~\cite{tomterud_observation_2025} as well as in the suspended form~\cite{blees_graphene_2015} from the literature. The $\kappa$'s of all 2D materials follow a universal, material-independent power-law scaling, $\kappa\left(L\right)\sim L^{0.82}$ for large $L$, consistent with the previous predictions of the power-law exponent in the range of 0.82-0.85 using the SCSA, the RG formulation and the atomistic calculations for 2D elastic membranes~\cite{le_doussal_self-consistent_1992, aronovitz_fluctuations_1988, los_scaling_2009, bowick_non-hookean_2017}. Furthermore, the $\kappa$'s asymptotically approach $\kappa_0\left(T\right)$ as $L\to 0$ for all 2D monolayers. The point of transition from the power-law regime for $\kappa\left(L\right)$ to a constant $\kappa_0$ occurs at different characteristic lengths $L_{\text{critical}} \sim 2\pi/q_{\text{critical}}$ for different materials. This material-dependence for the onset of such a regime transition can be inferred from the dimensionless vacuum polarization $\left[b_0\left(T\right)I\left(\mb{k}\right)\right]$, which causes the transition from small-$L$ $\left[b_0\left(T\right)I\left(\mb{k}\right)\ll 1\right]$ to large-$L$ $\left[b_0\left(T\right)I\left(\mb{k}\right)\gg 1\right]$ regime through its effect on the self-energy $\left[\Sigma\left(\mb{q}, T\right)\right]$~\cite{le_doussal_self-consistent_1992}. Due to the expected power-law scaling of the renormalized propagator $\left[G^{-1}\left(\mb{q}\right)\sim \frac{k_B T}{\kappa_0}q^{4-\eta_\kappa}\right]$, the dimensionless vacuum polarization can be separated into a product of a material-dependent part and a material-independent part as $b_0\left(T\right)I\left(\mb{k}\right) = k_B T\frac{Y^{\text{2D}}_0\left(T\right)}{2\kappa_0\left(T\right)^2}\int\frac{\mb{d}\mb{p}}{\left(2\pi\right)^2}\left[\mb{p}P^T\left(\mb{k}\right)\mb{p}\right]^2\tilde{G}\left(\mb{p}\right)\tilde{G}\left(\mb{k}-\mb{p}\right)$, where $\tilde{G}\left(\mb{p}\right) = \frac{\kappa_0\left(T\right)}{k_B T}G\left(\mb{q}\right)$ is material-independent. Thus, the material-dependent factor $\frac{Y^{\text{2D}}_0\left(T\right)}{2\kappa_0\left(T\right)^2}$ controls the difference in the critical wave vectors ($k = q_{\text{critical}}$) at which $b_0\left(T\right)I\left(\mb{k}\right)$ exceeds $1$ and the transition from $\kappa\to\kappa_0\left(T\right)$ to $\kappa\sim \kappa_0\left(T\right)L^{\eta_{\kappa}}$  happens, across different 2D monolayers. \\

\begin{figure}[!ht]
\centering
\includegraphics[width=\linewidth, trim=0mm 0mm 10mm 0mm, clip]{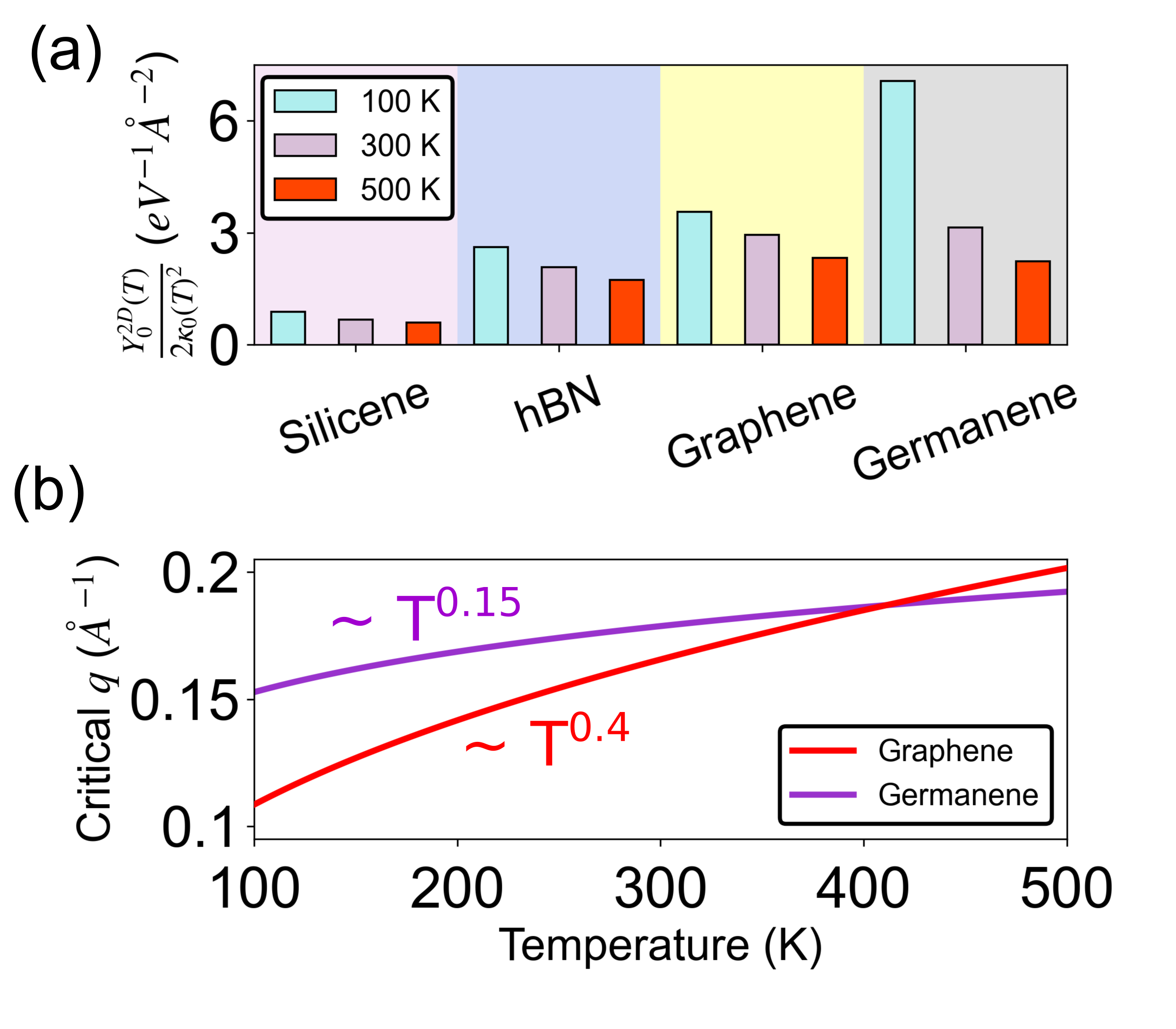}
\caption{\textbf{Temperature dependence of the critical wave vector ($q_{\text{critical}}$) for the renormalization of the elastic constants and its physical origin.} \textbf{(a)} The material-dependent factor $\frac{Y^{\text{2D}}_0\left(T\right)}{2\kappa_0\left(T\right)^2}$ at different temperatures for four 2D monolayers. \textbf{(b)} Temperature-dependence of $q_{\text{critical}}$ for graphene and germanene. The strong temperature-dependence of $\frac{Y^{\text{2D}}_0\left(T\right)}{2\kappa_0\left(T\right)^2}$ in germanene weakens the temperature-dependence of $q_{\text{critical}}$, resulting in a $\sim T^{0.15}$ scaling as opposed to a $\sim T^{0.4}$ scaling observed for graphene. Our predicted $q_{\text{critical}}$ for graphene at 300 K is consistent with that predicted using atomistic calculations~\cite{bowick_non-hookean_2017}.}  \label{fig:Figure6}
\end{figure}

As shown in Fig.~\ref{fig:Figure6} (a), the ratio $Y^{\text{2D}}_0\left(T\right)/\left[2\kappa_0\left(T\right)^2\right]$ is the smallest for silicene, followed by hBN, graphene and germanene at all temperatures. Hence, the critical wave vector ($k = q_{\text{critical}}$) at which the dimensionless vacuum polarization $\left[b_0\left(T\right)I\left(k\right)\right]$ exceeds $1$ in Fig.~\ref{fig:Figure3} (b) is the smallest for silicene and increases towards germanene in the same order as above. The ratio $Y^{\text{2D}}_0\left(T\right)/\left[2\kappa_0\left(T\right)^2\right]$, and therefore, $q_{\text{critical}}$, for the TMDCs are much smaller than that of silicene (see Supplementary Table II), hence they are not shown in Fig.~\ref{fig:Figure6} (a). Thus, the critical sample length $L\sim 2\pi/q_{\text{critical}}$ at which $\kappa\left(L\right)$ deviates from $\kappa_0\left(T\right)$ at 300 K in Fig.~\ref{fig:Figure3} (a) is the largest for MoS$_2$ and the smallest for germanene. In fact, even though the $\kappa_0\left(T\right)$ at 300 K of graphene, hBN and silicene are nearly identical to each other, as noted from the Supplementary Table II, the $\kappa$ of silicene exhibits a delayed deviation from $\kappa_0\left(T\right)$ compared to graphene and hBN with increasing $L$ in Fig.~\ref{fig:Figure3} (a), due to its relatively lower $Y^{\text{2D}}_0\left(T\right)/\left[2\kappa_0\left(T\right)^2\right]$. It is worth noting that the dimensionless vacuum polarization $\left[b_0\left(T\right)I\left(k\right)\right]$ almost exactly overlaps for graphene and germanene over the entire range of wave vectors at 300 K, as seen from Fig.~\ref{fig:Figure3} (b), which is consistent with the nearly identical values of $Y^{\text{2D}}_0\left(T\right)/\left(2\kappa_0\left(T\right)^2\right)$ at 300 K for these materials from Fig.~\ref{fig:Figure6} (a). Consequently, the scaling trends of $\kappa\left(L\right)$ for these two materials appear very similar in Fig.~\ref{fig:Figure3} (a), even though graphene is seven times stiffer for in-plane stretching and almost three times rigid for out-of-plane bending relative to germanene at 300 K (see Supplementary Table II). Thus, 2D monolayers with low $\kappa_0\left(T\right)$ and large $Y^{\text{2D}}_0\left(T\right)$ exhibit the strongest renormalization of $\kappa\left(L\right)$, and therefore, the ZA phonon dispersions as $q\to 0$.\\

We also obtain the renormalization of the 2D Young's modulus (as well as the 2D shear modulus and the Poisson's ratio - see Supplementary Fig. 6) from the dimensionless vacuum polarization within the SCSA framework as $Y^{\text{2D}}\left(L, T\right) = \frac{2b_0\left(T\right)k_B T}{1 + b_0\left(T\right)I\left(2\pi/L\right)}$. We find the same trends for $Y^{\text{2D}}\left(L, T\right)$ in Fig.~\ref{fig:Figure3} (c) as was observed for $\kappa\left(L, T\right)$ in Fig.~\ref{fig:Figure3} (a), i.e., $Y^{\text{2D}}\left(L, T\right)$ remains unchanged from $Y^{\text{2D}}_0\left(T\right)$ up to the critical sample length $L_{\text{critical}}$, after which it attains a universal power-law scaling $\left[Y^{\text{2D}}\left(L\right) \sim L^{-0.36}\right]$ for $L > 1\ \mu$m. The scaling exponent, $\eta_Y = -0.36$, is also consistent with the Ward identities for the rotation group in 2D, given by $\eta_Y = 2\eta_{\kappa} - 2$~\cite{le_doussal_self-consistent_1992, gazit_structure_2009}. Thus, the renormalization of $Y^{\text{2D}}\left(L\right)$ is much weaker than that of $\kappa\left(L\right)$. For example, while the $\kappa$ of a $~\sim 1\ \mu$m graphene sheet is about a hundred times larger than $\kappa_0\left(T\right)$ at 300 K, the $Y^{\text{2D}}$ is only about six times lower. Combining these observations with the weak renormalization observed for the 2D shear modulus and the Poisson's ratio in the Supplementary Fig. 6, we find that the renormalization of the ZA phonons, determined by $\kappa$, is much stronger than that of the in-plane LA and TA phonons, determined by $Y^{\text{2D}}$ and $\sigma$, in the long wavelength limit in large 2D monolayers. \\

\begin{figure*}[!ht]
\centering
\includegraphics[width=\linewidth, trim=4mm 0mm 13mm 0mm, clip]{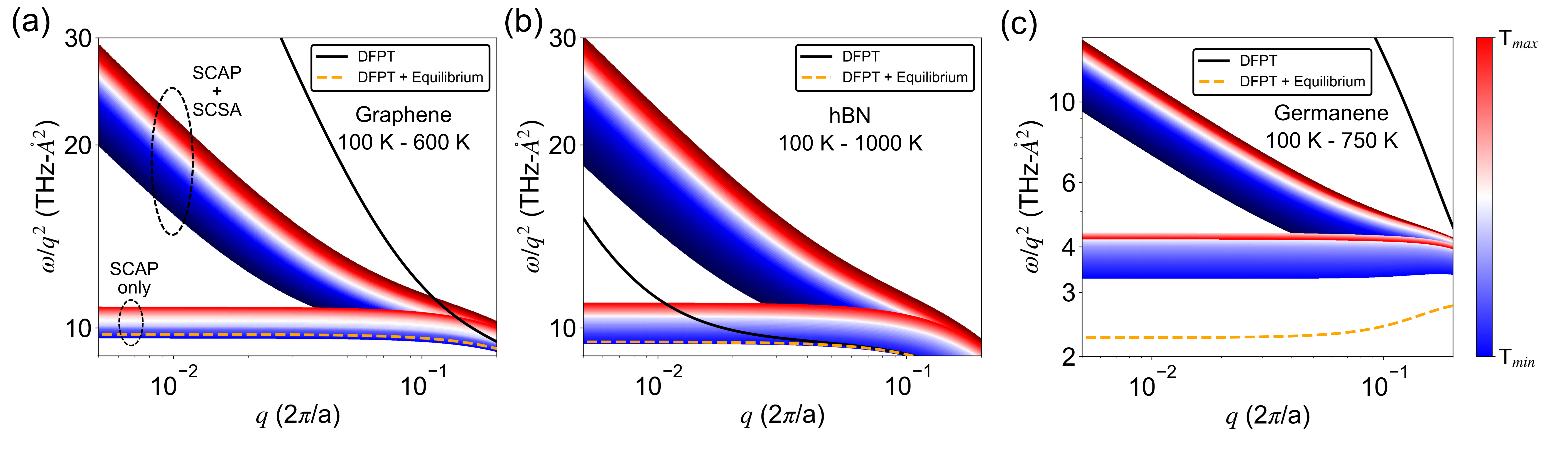}
\caption{\textbf{Renormalization of the ZA phonon dispersion in 2D monolayers.} \textbf{(a)} The ratio $\omega/q^2$ plotted as a function of $q$ for graphene in the temperature range of $\left[\text{T}_{min}, \text{T}_{max}\right]$ = $\left[100\ \text{K}, 600\ \text{K}\right]$. The solid black line is the result from the DFPT implementation in Quantum ESPRESSO, and the dashed yellow line is the result after the Born-Huang rotational invariance and stress-free equilibrium conditions are enforced on the DFPT result from Quantum ESPRESSO, representing the phonon modes obtained from the bare harmonic IFCs (i.e., $T \sim 0$ K). The colored contours are the results after the application of the SCAP renormalization described in this work. The curves plotted with the bright color-map represent the result of the SCAP implementation that are computed from the renormalized harmonic IFCs at different temperatures, while preserving the rotational invariance and stress-free equilibrium conditions. The curves plotted with the dark color-map are the results of the SCSA framework applied to the temperature-dependent phonons and the elastic constants from SCAP, thus capturing the departure from the quadratic behavior of the ZA phonon dispersion at small $q$. \textbf{(b)} Same as in \textbf{(a)} for hBN in the temperature range of $\left[\text{T}_{min}, \text{T}_{max}\right]$ = $\left[100\ \text{K}, 1000\ \text{K}\right]$. \textbf{(c)} Same as in \textbf{(a)} for germanene in the temperature range of $\left[\text{T}_{min}, \text{T}_{max}\right]$ = $\left[100\ \text{K}, 750\ \text{K}\right]$.}  \label{fig:Figure4}
\end{figure*}

These renormalized ZA dispersions are shown in Fig.~\ref{fig:Figure4} for graphene, hBN and germanene - the candidates exhibiting the strongest effect, along with their bare dispersions obtained from Quantum ESPRESSO~\cite{giannozzi_quantum_2009} using density functional perturbation theory (DFPT). The renormalization using the first-principles SCAP framework introduces the temperature dependence on the ZA phonon dispersions while enforcing the rotational invariance and stress-free equilibrium conditions on the renormalized IFCs to recover their quadratic dispersions in the small-$q$ elastic limit. The corresponding in-plane LA and TA phonon dispersions are presented in the Supplementary Figs. 4 and 5 respectively. By fitting these quadratic ZA and linear LA/TA phonon dispersions to the results of the continuum plate theory, we obtain $\kappa_0\left(T\right)$, $Y^{\text{2D}}_0\left(T\right)$ and $\sigma_0\left(T\right)$, which are subsequently fed into the SCSA calculations to predict the renormalized ZA phonon dispersions that stabilize the flat phase of the 2D monolayers at a given temperature. We find that the renormalization introduced by the SCSA has a large effect on the ZA phonon dispersions of all 2D monolayers shown in Fig.~\ref{fig:Figure4}, while that introduced by SCAP has the largest effect on the ZA phonons of germanene owing to its anomalously low $\kappa_0$ as discussed earlier. In fact, as shown in Fig.~\ref{fig:Figure4} (c), the temperature-dependence of the ZA phonons is the strongest at low temperatures for germanene, due to the rapid increase in the thermal occupation of the low-$\kappa_0$ ZA phonons. Once the Bose-Einstein distribution function reaches the classical regime for the low frequency ZA phonons with increasing temperature, the thermal occupation increases at a slower rate, thus slowing down the renormalization effect.

\section{Discussion}
\subsection{A case for revisiting phonon hydrodynamics using sub-quadratic ZA phonons}
The predicted strong renormalization of the long wavelength ZA phonons in the 2D monolayers has important implications for their thermal and electronic properties, as well as their maneuverability at the microscale. Previous computational studies have identified the ZA phonons with small wave vectors as the dominant heat carriers in 2D materials like graphene. In these materials, phonon-phonon scattering processes driven by the anharmonicity of the crystal lattice act to resist the heat flow~\cite{lindsay_phonon_2014, seol_two-dimensional_2010}, and recent reports have predicted unusually strong higher-order scattering among four phonons that limit the heat-carrying ability of these small-$q$ ZA phonons~\cite{han_thermal_2023}. Furthermore, at low temperatures, these four-phonon scattering processes among small-$q$ flexural phonons have been predicted to sharply weaken the unconventional hydrodynamic heat flow in graphene~\cite{tur-prats_high-order_2025, li_effects_2025}, while calculations involving only three-phonon scattering processes predicted strongly hydrodynamic heat flow behavior~\cite{lee_hydrodynamic_2015}. In all of these previous first-principles works, the dispersions of the ZA phonons were reported to vary quadratically with the phonon wave vector throughout the Brillouin zone. However, the renormalization of the small-$q$ ZA phonons that we predict at sample temperatures well above absolute zero and sample dimensions larger than a few 10's of nm could upend the above observations, since the number of allowed scattering channels satisfying the momentum and energy conservation requirements are very sensitive to the exact functional form of the phonon dispersions~\cite{ravichandran_phonon-phonon_2020}.

\subsection{Can 2D electrical resistivity logarithmically diverge when ZA phonons are renormalized?}
Apart from affecting thermal transport, ZA phonons also affect electronic transport in 2D materials. In 2D monolayers, a purely quadratic dispersion for the ZA phonons results in singularities in the electron mobility~\cite{fischetti_mermin-wagner_2016} and also causes a logarithmic divergence of the electrical resistivity with temperature, as discussed in Ref.~\cite{mariani_flexural_2008}. The latter study also arrived at a temperature-dependent critical wave vector $q_{\text{critical}} = \sqrt{\frac{3k_B T}{4\pi}}\sqrt{\frac{Y^{\text{2D}}_0}{2\kappa_0^2}}$ below which the ZA phonon dispersion gets renormalized as $\omega_{\text{ZA}}\sim q^{3/2}$, thereby quenching the low-temperature logarithmic divergence of the electrical resistivity. However, these conclusions were based on an approximate field-theoretical approach that effectively ignores the renormalization of $Y^{\text{2D}}$ with the system size and does not include the temperature-dependence of $\kappa_0$ and $Y^{\text{2D}}_0$. In fact, our first-principles approach, which captures the renormalization of $\kappa$ and $Y^{\text{2D}}$ simultaneously, predicts the small-$q$ behavior of ZA phonon dispersion to be of the form $\omega\sim q^{2-\frac{\eta_{\kappa}}{2}}$, with $\eta_{\kappa} = 0.82$. Furthermore, the temperature-dependence of $\kappa_0\left(T\right)$ and $Y^{\text{2D}}_0\left(T\right)$ results in a scaling of $q_{\text{critical}}\sim T^{0.4}$ for graphene but a much weaker $q_{\text{critical}}\sim T^{0.15}$ for germanene, as shown in Fig.~\ref{fig:Figure6} (b). These findings could result in dramatically different temperature-dependencies of the electrical resistivity for graphene and germanene, thus motivating a re-examination of electron-phonon scattering in 2D materials using the renormalized ZA phonon dispersions derived from first-principles in our work.

\begin{figure}[!ht]
\centering
\includegraphics[width=\linewidth, trim=0mm 0mm 10mm 0mm, clip]{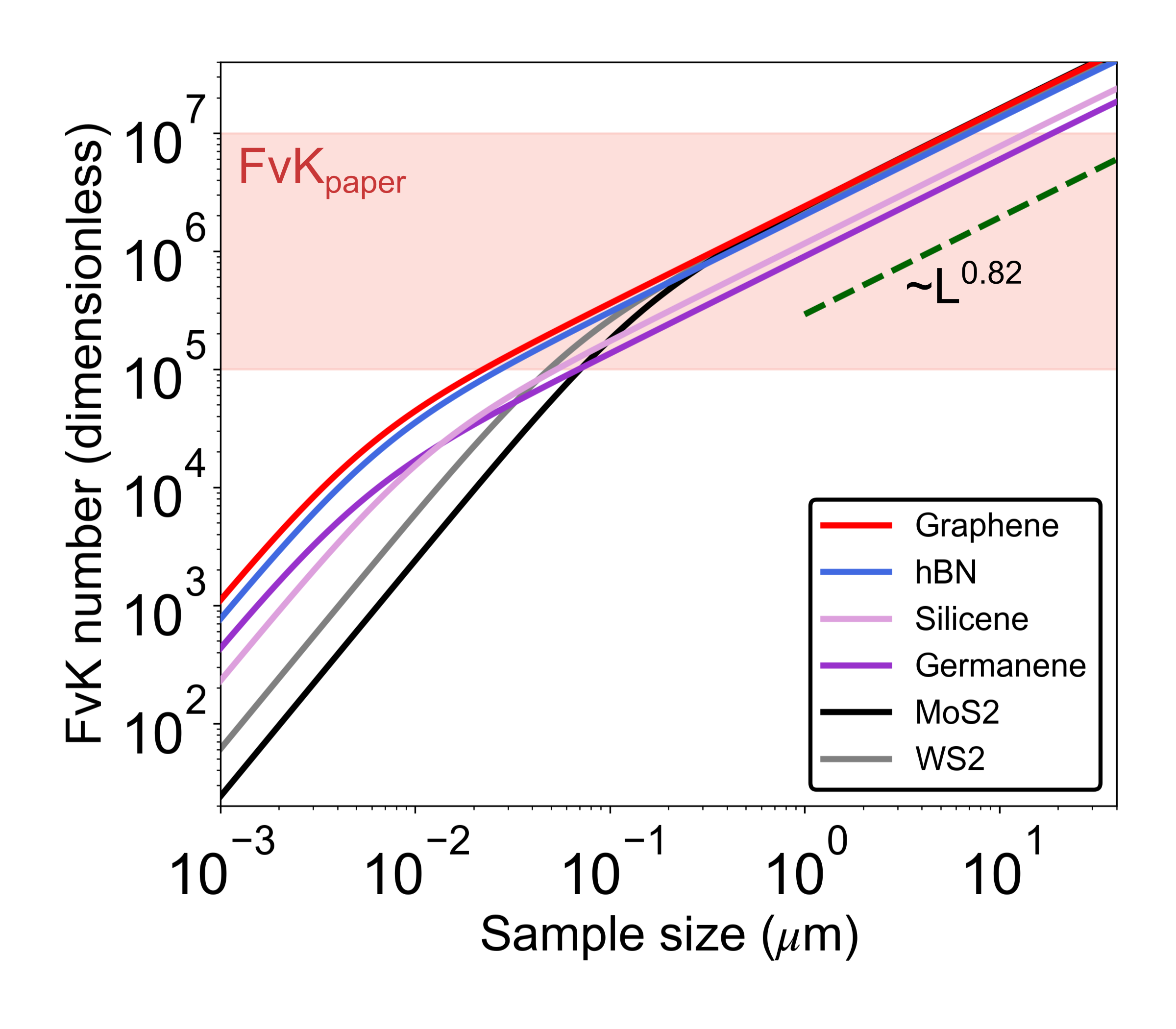}
\caption{\textbf{Effective Floppl von Karman (FvK) number for different 2D monolayers at 300 K.} The FvK shows four-to-six orders of magnitude enhancement when the sample size ($L$) is increased from $\sim 1$ nm to $\sim 40$ microns. FvK has a much stronger dependence on $L$ at the nanoscale, but achieves a universal scaling of FvK $\sim L^{0.82}$ for samples larger than a micron in size. All of the 2D monolayers have their FvK numbers in the range of that for a typical paper ($\sim 10^5-10^7$) for $L$ in the range of $\sim$ 100 nm to $\sim$ 10 $\mu$m, thus opening up possibilities for novel engineering applications such as kirigami beyond graphene~\cite{blees_graphene_2015}.}  \label{fig:Figure5}
\end{figure}

\subsection{Material alternatives to graphene for microscale kirigami}
Finally, our first-principles prediction of temperature- and size-dependent elastic constants will inform the choice of the nature of samples and experimental conditions under which novel engineering applications such as kirigami – the Japanese art of cutting and folding 2D paper to achieve intricate three-dimensional geometries, can be realized in micron-scale 2D materials. Drawing inspiration from conventional paper-kirigami, Blees et al.~\cite{blees_graphene_2015} realized kirigami in graphene by estimating the size of graphene sheet needed to match the non-dimensional effective Floppl von-Karman (FvK) number of a thin sheet of paper, defined as $Y^{\text{2D}}L^2/\kappa$. Using our predictions of the elastic constants, we predict the sample sizes where the effective FvK of different 2D monolayers lie within the range of FvK for regular paper in Fig.~\ref{fig:Figure5}. All 2D monolayers with sample sizes exceeding 100 nm up to about 10 $\mu$m exhibit FvK comparable to that of paper, thus encouraging development of kirigami applications in 2D materials beyond graphene.

\section{Summary}
In summary, we have demonstrated, using first-principles calculations, that the anharmonicity of the crystal potential at the nanoscale and the elastic properties at the macroscale jointly shape the phonon dispersions of 2D monolayers. Their effects are particularly strong on the out-of-plane flexural (ZA) phonons in materials with low bending rigidities ($\kappa$) and large Young's moduli $\left(Y^{\text{2D}}\right)$ at temperatures above absolute zero. The resulting renormalized ZA phonon dispersions show strong temperature dependence in low-$\kappa$ materials like germanene, due to the high thermal occupation of the ZA phonons in these materials. Furthermore, the dispersion of the ZA phonons exhibits a departure from a purely quadratic form into a sub-quadratic dispersion even for wavelengths as short as 10 nm in monolayers like graphene and germanene with large $Y^{\text{2D}}$ relative to $\kappa$, thus highlighting the importance of the renormalization effect even for samples as small as a few 10's of nanometers. We anticipate that our findings will motivate re-examination of the unconventional thermal and electronic phenomena predicted in the literature, that rely on the assumption of a temperature-independent, perfectly quadratic ZA phonon dispersion in 2D materials. The methodology introduced in Ref.~\cite{sarkar_statistical_2023} presents a direction to generalize our first-principles renormalization framework to other low-dimensional systems such as nanotubes and crystalline polymer strands as well, thus opening up possibilities for the computational exploration of novel energy transport phenomena in such materials. We also expect that our findings will inform future experimental efforts to realize exciting engineering applications exclusive to low dimensions, such as kirigami, in materials beyond graphene. 

\section{Acknowledgments}
This work was supported by the Core Research Grant (CRG) No. CRG/2022/009160, and the Mathematical Research Impact Centric Support (MATRICS) Grant No. MTR/2022/001043 from the Science and Engineering Research Board, India, and by the Advanced Research Grant (ARG) No. ANRF/ARG/2025/007160/ENS from the Anusandhan National Research Foundation, India.

\appendix
\section{Anharmonic renormalization of ZA phonons via self-consistent anharmonic phonon (SCAP) framework}  \label{sec:methods_SCAP}
We compute the anharmonic renormalization of the ZA phonons using the self-consistent anharmonic phonon (SCAP) framework~\cite{ravichandran_unified_2018}. First, we obtain the bare harmonic IFC tensors - $\chi_{\alpha\beta}^{\left(M\mu, N\nu\right)}$ and the bare phonon dispersions - $\omega_{\mb{q}j}$ using the density functional perturbation theory (DFPT), as implemented in Quantum ESPRESSO~\cite{giannozzi_quantum_2009}. Here, $M$ and $N$ are the indices for the lattice sites, $\mu$ and $\nu$ are those for the basis atoms, $\alpha$ and $\beta$ are the Cartesian components, $\mb{q}$ is the phonon wave vector and $j$ is the phonon polarization. As noted in Refs.~\cite{carrete_physically_2016, lin_general_2022}, the dispersion relation for the ZA phonons of unstrained 2D monolayers obtained from first principles calculations must have a quadratic dependence on $q$ in the small-$q$ limit as long as the appropriate crystal space group symmetries, invariance conditions and equilibrium requirements are enforced on the harmonic IFCs. While the IFCs obtained using the DFPT as implemented in Quantum ESPRESSO satisfies crystal point group symmetries and translational invariance conditions, we enforce the Born-Huang rotational invariance and the Huang condition for stress-free equilibrium requirements~\cite{croy_bending_2020, lin_general_2022} through in-house implementation. Additionally, for polar materials such as the TMDCs and hBN, we incorporate the long-range and non-analytical corrections to phonon frequencies, specifically developed for 2D systems, as described in Refs.~\cite{sohier_two-dimensional_2016, sohier_breakdown_2017}. The first-principles parameters that result in converged dispersion relations for the 2D monolayers in this study are tabulated in the Supplementary Table I. As shown in Fig.~\ref{fig:Figure1} (a), our calculated dispersion relations for the ZA phonons in the small-$q$ limit exhibit a quadratic relationship with $q$ for all 2D monolayers considered in this work. \\

Using these bare harmonic IFCs ($\chi_{\alpha\beta}^{\left(M\mu, N\nu\right)}$) from DFPT as a starting point, we obtain the temperature-dependent harmonic IFCs ($\psi_{\alpha\beta}^{\left(M\mu, N\nu\right)}$) and subsequently, the temperature-dependent phonon dispersions, using the self-consistent anharmonic phonon renormalization approach - SCAP~\cite{ravichandran_unified_2018}. In SCAP, the renormalized harmonic IFCs ($\psi_{\alpha\beta}^{\left(M\mu, N\nu\right)}$) including the effects of the quantum zero-point motion are computed by solving the following equation in a self-consistent manner:
\begin{align}
    & \psi_{\alpha\beta}^{\left(M\mu, N\nu\right)} = \chi_{\alpha\beta}^{\left(M\mu, N\nu\right)} + \frac{1}{2}\sum_{PQ, \pi\eta, \gamma\delta}\chi_{\alpha\beta\gamma\delta}^{\left(M\mu, N\nu, P\pi, Q\eta\right)}\nonumber\\
    &\ \ \ \ \ \ \ \ \ \ \ \ \ \ \times\expval{U_\gamma\left(P\pi\right)U_\delta\left(Q\eta\right)},\ \ \ \ \text{where}\nonumber\\
    &\expval{U_\gamma\left(P\pi\right)U_\delta\left(Q\eta\right)} = \frac{\hbar}{2N_0}\sum_{PQ, \pi\eta, \gamma\delta, \mb{q}j}\chi_{\alpha\beta\gamma\delta}^{\left(M\mu, N\nu, P\pi, Q\eta\right)}\nonumber\\
    &\ \ \ \ \ \times\frac{W_{\gamma}\left(\pi, \mb{q}j\right) W_{\delta}\left(\eta, -\mb{q}j\right)}{\Omega_{\mb{q}j}\sqrt{M_\pi M_\eta}}e^{i\mb{q}\cdot\left(R\left(P\pi\right)-R\left(Q\eta\right)\right)}\left(2n_{\mb{q}j} + 1\right)   \label{eq:self_consistent_renorm}
\end{align}

Here, $\chi_{\alpha\beta\gamma\delta}^{\left(M\mu, N\nu, P\pi, Q\eta\right)}$ are the bare quartic IFC tensors with the Cartesian indices $\left(\alpha, \beta, \gamma, \delta\right)$ and the atomic indices $\left(M\mu, N\nu, P\pi, Q\eta\right)$, $U_\gamma\left(P\pi\right)$ is the displacement of atom $\left(P\pi\right)$ from its equilibrium position along the Cartesian direction $\gamma$, and $\Omega_{\mb{q}j}$, $\mb{W}\left(\mb{q}j\right)$ and $n_{\mb{q}j}$ are the renormalized phonon frequencies, eigenvectors and Bose factors of the mode $\mb{q}j$ corresponding to the renormalized harmonic IFCs ($\psi_{\alpha\beta}^{\left(M\mu, N\nu\right)}$). Intuitively, the renormalization of the harmonic IFCs by the SCAP framework can be understood as follows - the bare harmonic IFCs obtained from DFPT correspond to the forces felt by the atoms of a crystal when they are displaced infinitesimally from their equilibrium configuration at $T \sim 0$ K. However, in reality, these atoms undergo thermal as well as quantum zero-point motion about their equilibrium configuration. These displacements result in large thermal excursions of atoms away from equilibrium that are appropriate for the given temperature $T>0$ K, thereby enabling the exploration of the anharmonic part of the potential energy surface of a real crystal. Furthermore, these thermal displacements also modify the average environmental configuration as seen by the atoms. Thus, the atoms are subject to the effective renormalized harmonic forces that are different from the bare forces obtained from DFPT, which is captured by Eq.~\ref{eq:self_consistent_renorm}.\\

We obtain the bare quartic IFCs ($\chi_{\alpha\beta\gamma\delta}^{\left(M\mu, N\nu, P\pi, Q\eta\right)}$) using the thermal snapshot technique, as described in Ref.~\cite{ravichandran_unified_2018}, and solve Eq.~\ref{eq:self_consistent_renorm} iteratively in a self-consistent manner. We enforce the crystal point group symmetries and translational invariance conditions by symbolically identifying the irreducible set of IFCs that satisfy these conditions, and solve Eq.~\ref{eq:self_consistent_renorm} only for the components of the harmonic IFCs belonging to that irreducible set. We enforce the Born-Huang rotational invariance condition and the Huang condition for stress-free equilibrium on the above-described irreducible set of IFCs using a Lagrange multiplier method at every iterative step while solving Eq.~\ref{eq:self_consistent_renorm}. Finally, the entire set of IFCs are obtained by back-propagation from the irreducible set using symbolic mapping and then, the renormalized frequencies ($\Omega_{\mb{q}j}$), eigenvectors ($\mb{W}\left(\mb{q}j\right)$) and the Bose factors ($n_{\mb{q}j}$) are obtained at each iterative step. Since the Born-Huang rotational invariance and the Huang stress-free equilibrium conditions are enforced during every iterative step while solving Eq.~\ref{eq:self_consistent_renorm}, and since the corrections to the ZA phonon frequencies at small $q$ due to SCAP (from Eq.~\ref{eq:self_consistent_renorm}) preserves the quadraticity of the ZA phonon dispersion for the 2D monolayers in this study, as shown in the Supplementary Section 3, the resulting renormalized ZA phonon dispersions continue to exhibit a quadratic dependence on $q$ in the small-$q$ limit. The temperature-dependent dispersion relations along the high symmetry directions of the BZ for the 2D monolayers considered in this study are shown in Supplementary Figs. 1-3.\\

\section{Renormalization of ZA phonons in the elastic limit via self-consistent screening approximation (SCSA)}  \label{sec:methods_SCSA}
In 2D materials, a perfectly quadratic ZA phonon dispersion forbids long-range ordering of the atoms in the lattice to stabilize the flat phase of large samples, as detailed in the main text. To stabilize the flat phase in 2D, the coupling between the in-plane and the out-of-plane degrees of freedom must be introduced in the free energy, which renormalizes the ZA phonon dispersion in the small-$q$ limit. We calculate this small-$q$ renormalization of the ZA phonons using the self-consistent screening approximation (SCSA) framework~\cite{le_doussal_self-consistent_1992, gazit_structure_2009, zakharchenko_self-consistent_2010, kosmrlj_mechanical_2013} with first-principles inputs of the temperature-dependent elastic constants obtained from the SCAP framework.\\

In micron-scale 2D monolayers, the out-of-plane displacements contribute to the stretching free energy, apart from the bending free energy introduced earlier, through the symmetric part of the strain tensor $u_{ij} = \frac{1}{2}\left(\partial_iu_j + \partial_ju_i + \partial_ih\partial_jh\right)$. Here, the strain tensor $u_{ij}$ contains the leading-order terms in the displacement gradients. In this case, the free energy of the 2D monolayers is given by:
\begin{align}
    F\left(\mb{u}, h\right) &= \frac{1}{2A}\int \left[\kappa\left(\nabla^2 h\right)^2 + 2\mu u_{ij}u_{ij} + \lambda u_{ii}^2\right]d\mb{x}
\end{align}
where $\left(\lambda, \mu\right)$ are the Lam\'e constants. Following Ref.~\cite{nelson_membraneStatMech_2004}, we integrate out the in-plane degrees of freedom, since they appear only quadratically in $F\left(\mb{u}, h\right)$, to obtain an effective free energy, $\tilde{F}\left(h\right)$, as an integral over in-plane Fourier variables $\left[\mb{q}, \mb{k}\right]$ as:
\begin{align}
    \tilde{F}\left(h\right) &= \frac{1}{2}\int \frac{\mathrm{d}\mb{q}}{4\pi^2}\kappa q^4\left|h\left(\mb{q}\right)\right|^2 \nonumber\\
    &\ \ + \frac{1}{2}\int \frac{\mathrm{d}\mb{q}}{4\pi^2}\int \frac{\mathrm{d}\mb{k}}{4\pi^2}\int\frac{\mathrm{d}\mb{k}'}{4\pi^2}\frac{S\left(\mb{k}, \mb{k}', \mb{q}\right)}{4}\nonumber\\
    &\ \ \ \ \ \ \ \ \ \ \times h\left(\mb{k}\right)h\left(\mb{q-k}\right)h\left(\mb{k}'\right)h\left(\mb{-q-k}'\right) \label{eq:int_FE}
\end{align}
where the four-point coupling is given by:
\begin{align}
    S\left(\mb{k}, \mb{k}', \mb{q}\right) &= 2\mu\left[\mb{k}^T\tilde{P}\left(\mb{q}\right)\mb{k}'\right]^2 \nonumber\\
    &\ \ \ \ + \frac{2\mu\lambda}{2\mu + \lambda}\left[\mb{k}^T\tilde{P}\left(\mb{q}\right)\mb{k}\right]\left[\left(\mb{k}'\right)^T\tilde{P}\left(\mb{q}\right)\mb{k}'\right]
\end{align}
with the elements of the transverse projection operator, $\tilde{P}_{\alpha\beta}\left(\mb{q}\right)$, given by:
\begin{align}
    \tilde{P}_{\alpha\beta}\left(\mb{q}\right) = \delta_{\alpha\beta} - \frac{q_\alpha q_\beta}{\|\mb{q}\|^2}
\end{align}
In two dimensions, we have the additional simplification that 
\begin{align}
    \left[\mb{k}P^{T}\left(\mb{p}\right)\mb{k}'\right]^2 =  \left[\mb{k}P^{T}\left(\mb{p}\right)\mb{k}\right]\left[\mb{k}'P^{T}\left(\mb{p}\right)\mb{k}'\right] = \left[\mb{k}\times\mb{\hat{p}}\right]^2 \left[\mb{k}'\times\mb{\hat{p}}\right]^2
\end{align}

From $\tilde{F}\left(h\right)$, we obtain the system size-dependent elastic constants of the 2D monolayers by solving the Dyson's equation for the renormalized propagator $\left[G^{-1}\left(\mb{q}, T\right)\right]$~\cite{gazit_structure_2009, kosmrlj_mechanical_2013}:
\begin{align}
    \frac{\kappa\left(\mb{q}, T\right)}{k_BT}q^4 = G^{-1}\left(\mb{q}, T\right) &= G_0^{-1}\left(\mb{q}, T\right) + \Sigma\left(\mb{q}, T\right) \label{eq:dyson}
\end{align}
in a self-consistent manner. Here, $G_0^{-1}\left(\mb{q}, T\right) = \frac{\kappa_0\left(T\right)}{k_B T}q^4$ is the bare propagator and the self energy, $\Sigma\left(\mb{q}, T\right)$, is given by:
\begin{align}
    \Sigma\left(\mb{q}, T\right) &= 2\int\frac{d^2\mb{k}}{\left(2\pi\right)^2}\frac{b_0\left(T\right)}{1 + b_0\left(T\right)I\left(\mb{k}\right)}\left[\mb{q}P^T\left(\mb{k}\right)\mb{q}\right]^2\nonumber\\
    &\ \ \ \ \ \ \ \ \ \ \ \ \ \ \ \ \ \ \ \ \ \ \ \times G\left(\mb{q}-\mb{k}\right)\nonumber\\
    &= 2\int\frac{d^2\mb{k}}{\left(2\pi\right)^2}\frac{b_0}{1 + b_0I\left(\mb{k}\right)}\|\hat{\mb{k}}\times\mb{q}\|_2^4G\left(\mb{q}-\mb{k}\right)
\end{align}
where $b_0\left(T\right) = Y^{\text{2D}}_0\left(T\right)/\left(2k_B T\right)$ and $I\left(\mb{k}\right) = \int\frac{\mb{d}\mb{p}}{\left(2\pi\right)^2}\left[\mb{p}P^T\left(\mb{k}\right)\mb{p}\right]^2G\left(\mb{p}\right)G\left(\mb{k}-\mb{p}\right)$ is related to the vacuum polarization integral introduced in Ref.~\cite{le_doussal_self-consistent_1992}.\\

We obtain the elastic constants as functions of the system size ($L\approx 2\pi/q$) by solving Eq.~\ref{eq:dyson} for the renormalized propagator $G^{-1}\left(\mb{q}, T\right)$ in a self-consistent manner to get the Fourier-domain variables $\kappa\left(\mb{q}, T\right)$, $Y^{\text{2D}}\left(\mb{q}, T\right) = \frac{2b_0\left(T\right)k_B T}{1 + b_0\left(T\right)I\left(\mb{q}\right)}$, the shear modulus $\mu\left(\mb{q}, T\right) = \frac{\mu_0\left(T\right)}{1 + \frac{2}{3}\mu_0\left(T\right)I\left(\mb{q}\right)}$ and the Poisson's ratio $\sigma\left(\mb{q}, T\right) = \frac{Y^{\text{2D}}\left(\mb{q}, T\right)}{2\mu\left(\mb{q}, T\right)} - 1$. Subsequently, we obtain the renormalized ZA, LA and TA phonon frequencies in the small-$q$ limit using $\kappa\left(\mb{q}, T\right)$, $Y^{\text{2D}}\left(\mb{q}, T\right)$ and $\sigma\left(\mb{q}, T\right)$ in the parametrized dispersion relation introduced in the main text. \\

In this work, we solve Eq.~\ref{eq:dyson} numerically in 2D polar coordinates, with the self-energy and the vacuum polarization integrals given by:
\begin{align}
    \Sigma&\left(q, \phi\right) = \frac{1}{2\pi^2}\int_0^{2\pi}\mathrm{d}\psi\int_0^{q_{\text{max}}}\frac{b_0}{1 + b_0I\left(k, \psi\right)}kq^4\sin^4\left(\phi-\psi\right)\nonumber\\
    &\ \ \ \ \ \ \times G\left(\sqrt{q^2 + k^2 - 2qk\cos\left(\phi-\psi\right)}, \phi-\psi\right)\mathrm{d}k    \label{eq:self_energy}\\
    I&\left(q, \phi\right) = \frac{1}{4\pi^2}\int_{0}^{2\pi}\mathrm{d}\psi\int_{0}^{q_{\text{max}}}k^5\sin^4\left(\phi-\psi\right)\nonumber\\
    &\ \ \times G\left(k, \psi\right)G\left(\sqrt{q^2 + k^2 - 2qk\cos\left(\phi-\psi\right)}, \phi-\psi\right)\mathrm{d}k   \label{eq:vacuum_pol}
\end{align}

While solving Eqs.~\ref{eq:dyson}, ~\ref{eq:self_energy} and~\ref{eq:vacuum_pol} self-consistently, we use two-dimensional Gauss quadrature to approximate the integrals. We evaluate the propagators with arbitrary arguments \Bigg[e.g., $G^{-1}\left(\sqrt{q^2 + k^2 - 2qk\cos\left(\phi-\psi\right)}, \phi-\psi\right)$\Bigg] by interpolating from the corresponding values on the quadrature points. By following the SCSA formulation as in Ref.~\cite{gazit_structure_2009, kosmrlj_mechanical_2013}, we avoid the numerical singularities in solving Eq.~\ref{eq:dyson} that was observed in Ref.~\cite{zakharchenko_self-consistent_2010}. We note that, since all 2D monolayers in this study have hexagonal symmetry, all propagators, self-energies and elastic properties are isotropic in the small-$q$ limit~\cite{burmistrov_emergent_2022}; hence they depend only on the magnitude of $\mb{q}$.\\
% \section{Code availability}
% All formulations and algorithms necessary to reproduce the findings of this work are
% presented in the Methods section and in Ref.~\cite{ravichandran_unified_2018}
% \section{Data availability}
% The data supporting the conclusions of this study will be made available by the corresponding author upon reasonable request.
\bibliographystyle{unsrt}
\bibliography{references_prx}
\end{document}